\documentclass[journal,twoside]{IEEEtran}
\usepackage{cite}
\usepackage{amsmath,amssymb,amsfonts, mathtools}
\usepackage{graphicx}
\usepackage{textcomp}
\usepackage{xcolor}
\usepackage{comment}
\usepackage{commath}
\usepackage{algorithm,algorithmic}
\usepackage{enumerate}
\usepackage{psfrag,subfigure}
\usepackage{amsbsy,epsfig,bbm,mathrsfs,multirow,amsthm}
\usepackage{float}
\usepackage{mathrsfs}
\usepackage{color}
\usepackage{tabulary}
\usepackage{multirow}
\usepackage{cases}
\usepackage{stfloats}
\usepackage{url}
\usepackage[subfigure]{tocloft}
\usepackage{import}
\usepackage{xcolor}

\DeclareMathOperator{\diag}{\mathrm{diag}}
\DeclareMathOperator*{\argmax}{arg\,max}
\DeclareMathOperator*{\clip}{\mathrm{clip}}

\begin{document}

\title{Deep Reinforcement Learning for Intelligent Reflecting Surface-assisted D2D Communications}

\author{Khoi Khac Nguyen, Antonino Masaracchia, Cheng Yin, Long D. Nguyen, Octavia A. Dobre, and Trung Q. Duong
\thanks{K. K. Nguyen, A. Masaracchia, C. Yin, and T. Q. Duong are Queen's University Belfast, UK (e-mail: \{knguyen02,a.masaracchia,cyin01,trung.q.duong\}@qub.ac.uk). L. D. Nguyen is with Duy Tan University, Vietnam (email: dinhlonghcmut@gmail.com). O. A. Dobre is with Memorial University, Canada (e-mail: odobre@mun.ca)}
}

\maketitle

\begin{abstract}
In this paper, we propose a deep reinforcement learning (DRL) approach for solving the optimisation problem of the network's sum-rate in device-to-device (D2D) communications supported by an intelligent reflecting surface (IRS). The IRS is deployed to mitigate the interference and enhance the signal between the D2D transmitter and the associated D2D receiver. Our objective is to jointly optimise the transmit power at the D2D transmitter and the phase shift matrix at the IRS to maximise the network sum-rate. We formulate a Markov decision process and then propose the proximal policy optimisation for solving the maximisation game. Simulation results show impressive performance in terms of the achievable rate and processing time.
\end{abstract}
\begin{IEEEkeywords}
Intelligent reflecting surface (IRS), D2D communications, deep reinforcement learning.
\end{IEEEkeywords}


\section{Introduction}\label{Sec:Intro}
Device-to-device (D2D) communications play a critical role in 5G networks by allowing users to communicate directly without the involvement of base stations. It helps reduce the latency and improve the information transmission efficiency \cite{Khoi:19:Access, JH:20:WC}. In \cite{Khoi:19:Access}, the optimised power allocation at the D2D transmitters was proposed to maximise the energy efficiency (EE) performance, by following a machine learning-based approach. In \cite{JH:20:WC}, the D2D transmitters harvest energy through the simultaneous wireless information and power transfer protocol (SWIPT). Then, a game theory approach was proposed to solve the power allocation and power splitting at SWIPT with pricing strategies for maximising the network performance.

Intelligent reflecting surface (IRS), referring to the technology of massive elements of flexible reflection capability that are controlled by an intelligent unit, has recently attracted great attention from the research community as an efficient means to expand wireless coverage. The IRS can manage the incoming signal by a controller, which allows to efficiently adapt the angle of passive reflection from the transmitters toward the receivers \cite{HY:20:JSAC, YZ:20:VT, BZ:21:TCOM, KK:21:NCE}. In \cite{YZ:20:VT}, the IRS harvests energy from the access point (AP) and uses it for reflecting the signal in two phases. The AP beamforming vector, the IRS's phase scheduling, and the passive beamforming were optimised to maximise the information rate. In \cite{BZ:21:TCOM}, a channel estimation scheme for a multi-user multiple-input multiple-output (MIMO) system has been designed with the support of double IRS panels.

Some research works have investigated the efficiency of the IRS in assisting the D2D communications \cite{YC:21:WC, SJ:21:WCL}. In \cite{YC:21:WC} and \cite{SJ:21:WCL}, two sub-problems with fixed passive beamforming vector and fixed phase shift matrix were considered. To solve the power allocation optimisation with the fixed phase shift matrix, the authors in \cite{YC:21:WC} used the gradient descent method while the authors in \cite{SJ:21:WCL} employed the Dinkelbach method. For the phase shift optimisation, a local search algorithm was proposed in \cite{YC:21:WC} while fractional programming was utilised in \cite{SJ:21:WCL}. However, these approaches assume a discrete phase shift and only reach a sub-optimal solution. Moreover, these works only consider perfect conditions, e.g., channel state information (CSI). In addition, these algorithms cause large delays due to high computational complexity.

Very recently, deep reinforcement learning (DRL) has been applied as an effective solution for solving complicated problems in wireless networks \cite{KK:19:Access, Khoi:20:Access, CH:20:JSAC, MS:21:VT, KF:20:WCL,  KK:21:TCOM}. In \cite{KK:19:Access}, we defined the discrete power level and used the DRL algorithm to choose the transmit power at the D2D transmitter for maximising the EE. In \cite{CH:20:JSAC}, discrete and continuous action spaces were considered for the beamforming vector and the IRS phase shift in multiple-input single-output (MISO) communications. Then, two DRL algorithms were used to maximise the total throughput. In \cite{MS:21:VT}, a method based on the DRL was used for optimising the unmanned aerial vehicle (UAV)'s altitude and the IRS diagonal matrix to minimise the sum age-of-information. In \cite{KF:20:WCL}, the authors used the DRL technique to maximise the signal-to-noise ratio.

In this paper, we propose a DRL algorithm for solving the joint power allocation and phase shift matrix optimisation in the IRS-assisted D2D communications. Firstly, we conceive a D2D communication system with the support of the IRS. The D2D channel is a combination of the direct link and the reflective link. The IRS is used for mitigating the interference and enhancing the information transmission channel. Secondly, we formulate a Markov decision process (MDP) \cite{BD:95:Book:v1} for the network throughput maximisation in the IRS-assisted D2D communications, in which the optimisation variables are the power at the D2D users and the phase shifts at the IRS. Then, a DRL algorithm is used to search for an optimal policy for maximising the network sum-rate. Finally, we compare the efficiency of our proposed methods with other schemes in terms of the achievable network sum-rate. 


\section{System Model and Problem Formulation}\label{Sec:Model}
\begin{figure}[h!]
	\centering
	\subfigure{\includegraphics[width=0.455\textwidth]{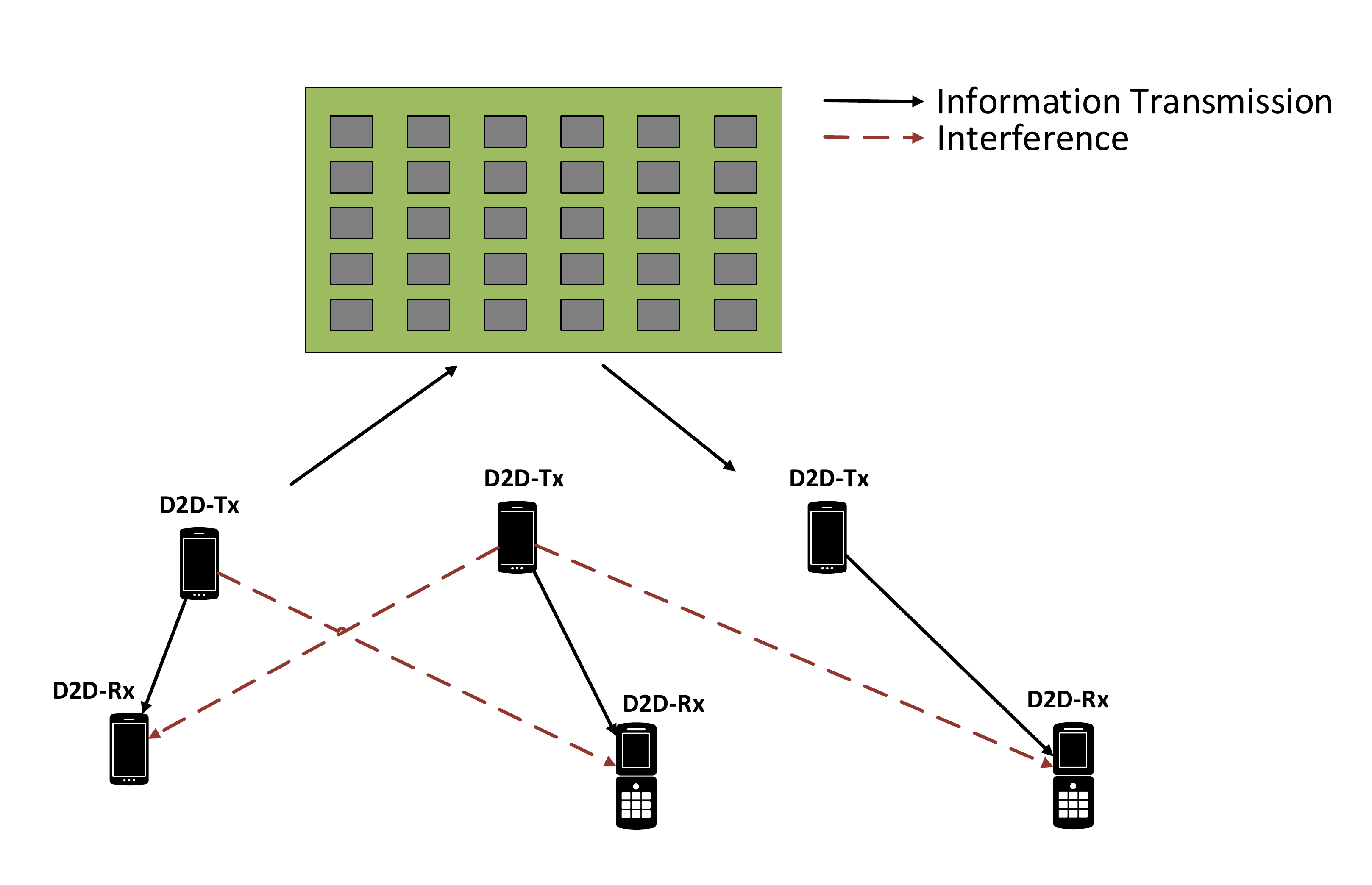}}
	\caption{System model of the IRS-assisted D2D communications.}
	\label{fig:System}
\end{figure}

We consider an IRS-assisted wireless network with $N$ pairs of D2D users distributed randomly and an IRS panel, as shown in Fig.~\ref{fig:System}. Each pair of D2D users comprises of a single-antenna D2D transmitter (D2D-Tx) and a single-antenna D2D receiver (D2D-Rx). An IRS panel with $K$ reflective elements is deployed to enhance the signal from the D2D-Tx to the associated D2D-Rx and mitigate the interference from other D2D-Txs. The IRS with reflective elements maps the receiver's signal by the value of the phase shift matrix controlled by an intelligent unit. The received signal at the D2D-Rx is composed of a direct signal and a reflective one.

We denote the position of the $n$th D2D-Tx at time step $t$ as $X^{t}_n(\text{Tx}) = \big(x^{t}_n (\text{Tx}), y^{t}_n (\text{Tx})\big), n = 1, \dots, N$ and that of the $m$th D2D-Rx as $X^{t}_m(\text{Rx}) = \big(x^{t}_m(\text{Rx})),  y^{t}_m (\text{Rx})\big), m = 1, \dots, N$. The IRS is fixed at the position $(x^t_{\text{IRS}}, y^t_{\text{IRS}}, z^t_{\text{IRS}})$. The phase shift value of each element in the IRS belongs to $[0, 2\pi]$.

We denote the direct channel from the $n$th D2D-Tx to the $m$th D2D-Rx at time step $t$ by $h_{nm}^t$, and the reflective channel by $H^t_{nm}$. The phase shift matrix at the IRS at time step $t$ is defined by $\Phi^t = \diag(\eta_2^t \theta_1^t, \eta_2^t \theta_2^t, \dots, \eta_K^t \theta_K^t)$, where $\eta_k^t \in [ 0, 1 ]$ and $\theta_k^t \in [0, 2\pi]$ represent the amplitude and the phase shift value, respectively. In this paper, we assume that the amplitudes of all elements are set to $\eta_k^t = 1$.

The distance between the $n$th D2D-Tx and the $m$th D2D-Rx at time step $t$ is defined as
\begin{equation}
d_{nm}^t = \sqrt{\big(x^{t}_n(\text{Tx}) - x^{t}_m(\text{Rx})\big)^2 + \big(y^{t}_n (\text{Tx})- y^{t}_m(\text{Rx})\big)^2}.
\end{equation}

Similarly, the distance between the $n$th D2D-Tx and the IRS is $d_{n, IRS}^t$ and the distance between the IRS and the $m$th D2D-Rx is $d_{IRS, m}^t$ at time step $t$. The direct channel is formulated as
\begin{equation}
h_{nm}^t = \hat{h}_n \sqrt{\beta_0  (d^{t}_{nm})^{-\kappa_0}},
\end{equation}
where $\beta_0$ and $\hat{h}_n$ are the channel power gain at the reference distance $d_0 = 1$ m and $\kappa_0$ is the path-loss exponent in the D2D link. Here, we assume that the small-scale fading follows the Nakagami-$m$ distribution with $m$ as the fading severity parameter.

The reflective channel via the IRS from the $n$th D2D-Tx toward the $m$th D2D-Rx is considered as a Rician fading channel at time step $t$ described by
\begin{equation}
H_{nm}^t = \sqrt{ \frac{\beta_1}{1+\beta_1}} \tilde{h}^{LoS}_{nm} + \sqrt{\frac{1}{\beta+1}} \tilde{h}^{NLoS}_{nm},
\end{equation}
where $\beta_1$ is the Rician factor, and $\tilde{h}^{LoS}_{nm}$, $\tilde{h}^{NLoS}_{nm}$ are the line-of-sight (LoS) and the non-line-of-sight (NLoS) components for the reflected channel, respectively. Specifically, the LoS component is defined as \cite{YC:21:WC}
\begin{equation}
\tilde{h}^{LoS}_{nm} = \sqrt{\beta_0 (d_{n, IRS}^t  d_{IRS, m}^t) ^{-\kappa_0}} e^{-j \theta'},
\end{equation}
 where $\theta' \in [0, 2 \pi]$ is the random phase. The NLoS component is defined as
 \begin{equation}
 \tilde{h}^{NLoS}_{nm} =  \sqrt{\beta_0 (d_{n, IRS}^t  d_{IRS, m}^t) ^{-\kappa_1}} \hat{h}^{NLoS}_{nm},
 \end{equation}
 where $\kappa_1$ is the path loss exponent for the NLoS component and the small-scale fading $\hat{h}^{NLoS}_{nm} \sim \mathcal{C} \mathcal{N} (0, 1)$ is i.i.d. complex Gaussian distribution with zero mean and unit variance.


The received signal at the $n$th D2D-Rx at time step $t$ can be written as
\begin{equation}
\begin{split}
s_n^t =  \Big(h_{nn}^t &+ \sum^K_{k=1} H_{nn}^t \Phi^t\Big)  \sqrt{p_n^t}u_n^t \\ &+ \sum_{m \ne n}^N  \Big(h_{mn}^t + \sum^K_{k=1} H_{mn}^t \Phi^t \Big)\sqrt{p_m^t}u_m^t + \varpi,
\end{split}
\end{equation}
where $p_n^t$ is the transmit power at the $n$th D2D-Tx at time step $t$, $u_n^t$ is the transmitted symbol from the $n$th D2D-Tx, and $\varpi \sim \mathcal{N} (0, \alpha^2)$ is the complex additive white Gaussian noise.

Accordingly, the received signal-to-interference-plus-noise ratio (SINR) at the $n$th D2D-Rx can be represented as
\begin{equation}
\gamma_n^t = \frac{| h_{nn}^t + \sum^K_{k=1} H_{nn}^t \Phi^t|^2 p_n^t}{ \sum_{m \ne n, m \in N}  |h_{mn}^t + \sum^K_{k=1} H_{mn}^t \Phi^t|^2 p_m^t + \alpha^2}.
\end{equation}

The achievable sum-rate at the $n$th D2D pair during time step $t$ is defined as
\begin{equation}
R_n^t = B \log_2(1+\gamma_n^t),
\end{equation}
where $B$ is the bandwidth. 


In this paper, we aim at optimising the power allocation of all $N$ pairs of D2D users $P = \{ p_1, p_2, \dots, p_N \}$ and the phase shift matrix $\Phi$ of the IRS to maximise the network sum-rate while satisfying all the constraints. The considered network optimisation can be formulated as follows:
\begin{equation}\label{equ:Prob}
\begin{split}
\max_{P, \Phi} \quad & R_{total}^t = \sum_{n=1}^N R_n^t\\
s.t. \quad & 0 < p_n < P_{\max}, \forall n \in N \\
& R_n^t \ge r_{\min}, \forall n \in N \\
& \theta_k \in [0, 2 \pi], \forall k \in K,\\
\end{split}
\end{equation}
where $P_{\max}$ is the maximum transmit power at the D2D-Tx and the constraint $R_n^t \ge r_{\min}, \forall n \in N$ indicates the quality-of-service (QoS) of the D2D communications.

\section{Joint Optimisation of Power Allocation and Phase Shift Matrix}\label{Sec:PPO}
Given the optimisation problem (\ref{equ:Prob}), we formulate the MDP with the agent, the state space $\mathcal{S}$, the action space $\mathcal{A}$, the transition probability $\mathcal{P}$, the reward function $\mathcal{R}$ and the discount factor $\zeta$. Let us denote $\mathcal{P}_{ss'}(a)$ as the probability when the agent takes action $a^t \in \mathcal{A}$ at the state $s = s^t \in \mathcal{S} $ and transfers to the next state $s' =s^{t+1} \in \mathcal{S}$. In particular, we formulate the MDP game as follows:
\begin{itemize}
	\item{\emph{State space}}: The channel gain of the D2D users forms the state space as
\begin{align}\label{equ:state}
	\mathcal{S} &= \Big\{h_{11} + \sum_{k=1}^K H_{11} \Phi, \dots, h_{1N} + \sum_{k=1}^K H_{1N} \Phi, \dots,
    \nonumber \\
    &h_{nm}+ \sum_{k=1}^K H_{nm} \Phi, \dots, h_{nN}+ \sum_{k=1}^K H_{nN} \Phi, \dots,
    \nonumber \\
    &h_{N1} + \sum_{k=1}^K  H_{N1} \Phi, \dots,  h_{NN} + \sum_{k=1}^K  H_{NN} \Phi  \Big\}.
\end{align}
	\item{\emph{Action space}}: The D2D-Txs adjust the transmit power and the IRS changes the phase shift for maximising the expected reward. Thus, The action space for the D2D users and the IRS is considered as follows:
	\begin{equation}
	\mathcal{A} = \{p_1, p_2, \dots, p_N, \theta_1, \theta_2, \dots, \theta_K \}.
	\end{equation}
	\item{\emph{Reward function}}: The agent needs to find an optimal policy for maximising the reward. In our problem, our objective is to maximise the network sum-rate; thus, the reward function is defined as $\mathcal{R} =$
	\begin{equation}\label{equ:reward}
	 \sum_{n=1}^N B \log_2 \Bigg(1 + \frac{| h_{nn} + \sum^K_{k=1} H_{nn} \Phi|^2 p_n}{ \sum_{m \ne n}^N  |h_{mn} + \sum^K_{k=1} H_{mn} \Phi|^2 p_m + \alpha^2}\Bigg).
	\end{equation}
\end{itemize}

By following the MDP, the agent interacts with the environment and receives the response to achieve the best expected reward. Particularly, the state of the agent at time step $t$ is $s^t$. The agent chooses and executes the action $a^t$ under the policy $\pi$. The environment responds with the reward $r^t$. After taking the action $a^t$, the agent moves to the new state $s^{t+1}$ with probability $P_{ss'}(a)$. The interactions are iteratively executed and the policy is updated for the optimal reward.


In this paper, we propose a DRL approach to search for an optimal policy for maximising the reward value in (\ref{equ:reward}). The optimal policy can be obtained by modifying the estimation of the value function or directly by the objective. We use an on-policy algorithm for our work, namely proximal policy optimisation (PPO) with the clipping surrogate technique \cite{JS:17:PPO}. Consider the probability ratio of the current policy and obtained policy $p^t_\theta = \frac{\pi(s, a; \theta)}{\pi (s, a; \theta_{old})}$, we need to find the optimal policy to maximise the total expected reward as follows:
\begin{equation}
\mathcal{L} (s, a; \theta)\! =\! \mathbb{E}\! \Bigg[\! \frac{\pi(s, a; \theta)}{\pi (s, a; \theta_{old})} A^\pi(s, a) \Bigg]
\!=\!\mathbb{E}\! \Bigg[\!  p^t_\theta A^\pi(s, a) \!\Bigg],
\end{equation}
where $\mathbb{E} [\cdot] $ is the expectation operation and $A^\pi (s, a) = Q^\pi(s,a) - V^\pi(s)$ denotes the advantage function \cite{JS:16:ICLR}; $V^\pi(s)$ denotes the state-value function while $Q^\pi(s,a)$ is the action-value function.

In the PPO method, we limit the current policy such that it does not go far from the obtained policy by using different techniques, e.g., the clipping technique and Kullback-Leiber \cite{JS:16:ICLR}. In this work, we use the clipping surrogate method to prevent the excessive modification of the objective value, as follows:
\begin{equation}
\begin{split}
\mathcal{L}^{\clip} (s, a; \theta) = \mathbb{E} \Bigg[ & \min \Big(p^t_\theta A^\pi(s, a), \\& \clip(p^t_\theta, 1-\epsilon, 1+\epsilon )A^\pi (s, a)\Big) \Bigg],
\end{split}
\end{equation}
where $\epsilon$ is a hyperparameter.

When the advantage $A^\pi(s,a)$ is positive, the term $(1+\epsilon)$ takes action. Meanwhile, for the negative case of the advantage $A^\pi(s,a)$, the term $(1-\epsilon)$ sets a ceiling to limit the objective value. Moreover, for the advantage function $A^\pi(s, a)$, we use \cite{Mnih:16}:
\begin{equation}
\label{equ:A}
A^\pi(s, a) = r^t + \zeta V^\pi(s^{t+1}) -V^\pi(s^t),
\end{equation}
where the state-value function $V^\pi(s)$ is obtained at the state $s$ under the policy $\pi$ as follows:
\begin{equation}
V^\pi = \mathbb{E} \Big \{\mathcal{R}|s, \pi \Big\}.
\end{equation}

To train the policy network, we store the transition into a mini-batch memory $D$ and then use the stochastic policy gradient (SGD) method to maximise the objective. By denoting the policy parameter by $\theta$, it is updated as
\begin{equation}
\label{equ:policyPPO}
\theta^{d+1} = \argmax \mathbb{E} \Big[\mathcal{L}(s, a ; \theta^d)\Big].
\end{equation}


The PPO algorithm for joint optimisation of the transmit power and the phase shift matrix in the IRS-aided D2D communications is presented in Algorithm~\ref{alg:PPO}, where $M$ denotes the maximum number of episodes and $T$ is the number of iterations during a period of time.

\begin{algorithm}[t!]
	\caption{Proposed approach based on the PPO algorithm for the IRS-assisted D2D communications.}
	\begin{algorithmic}[1]
		\label{alg:PPO}
		\STATE Initialise the policy $\pi$ with the parameter $\theta_\pi$
		\STATE Initialise other parameters
		\FOR{episode = $1,\dots, M$}
		\STATE Receive initial observation state $s^0$
		\FOR{iteration = $1,\dots, T$}
		\STATE Obtain the action $a^t$ at state $s^t$ by following the current policy
		\STATE Execute the action $a^t$
		\STATE Receive the reward $r^t$ according to (\ref{equ:reward})
		\STATE Observe the new state $s^{t+1}$
		\STATE Update the state $s^t = s^{t+1}$
		\STATE Collect set of partial trajectories with $D$ transitions
		\STATE Estimate the advantage function according to (\ref{equ:A})
		\ENDFOR
		\STATE Update policy parameters using SGD with mini-batch $D$
		\begin{equation}
		\theta^{t+1} = \argmax \frac{1}{D} \sum^{D} \mathcal{L}^{\clip}(s, a ; \theta^t)
		\end{equation}
		\ENDFOR
	\end{algorithmic}
\end{algorithm}

\section{Simulation Results}\label{Sec:Results}
For numerical results, we use Tensorflow 1.13.1 \cite{Abadi:16}. The IRS is deployed at $(0,0,0)$, while the D2D devices are randomly distributed within a circle of $100$ m from the center. The maximum distance between the D2D-Tx and the associated D2D-Rx is set to $10$ m. We assume $d/\lambda = 1/2$, and set the learning rate for the PPO algorithm to $0.0001$. For the neural networks, we initialise two hidden layers with $128$ and $64$ units, respectively. All other parameters are provided in Table \ref{tab:Params}. We consider the following algorithms in the numerical results.
\begin{itemize}
	\item \textbf{The proposed algorithm}: We use the PPO algorithm with the clipping surrogate technique to solve the joint optimisation of the power allocation and the phase shift matrix of the IRS.
	\item \textbf{Maximum power transmission (MPT)}: The D2D-Tx transmits information with maximum power, $P_{\max}$. We use the PPO algorithm to optimise the phase shift matrix of the IRS panel.
	\item \textbf{Random phase shift matrix selection (RPS)}: We optimise the power allocation at the D2D-Tx with random selection of the phase shift matrix $\Phi$.
	\item \textbf{Without IRS}: The D2D-Tx transmits information without the support of the IRS. We optimise the power allocation by using the PPO algorithm.
\end{itemize}

\begin{table}[h!]
	\renewcommand{\arraystretch}{1.2}
	\caption{SIMULATION PARAMETERS.}
	\label{tab:Params}
	\centering
	\begin{tabular}{l|l}
		\hline
		Parameters & Value \\
		\hline
		Bandwidth ($W$)  & $1$ MHz \\
		Path-loss parameter & $\kappa_0 = 2.5, \kappa_1 = 3.6$\\
		Channel power gain & $-30$ dB\\
		Rician factor & $\beta_1 = 4$\\
		Noise power & $\alpha^2 = -80$ dBm\\
		Clipping parameter & $\epsilon = 0.2$\\
		Discounting factor & $\zeta = 0.9$\\
		Max number of D2D pairs & $10$\\
		Initial batch size & $ K = 128$ \\
		
		\hline
	\end{tabular}
\end{table}

\begin{figure}[h!]
	\centering
	\subfigure{\includegraphics[width=0.45\textwidth]{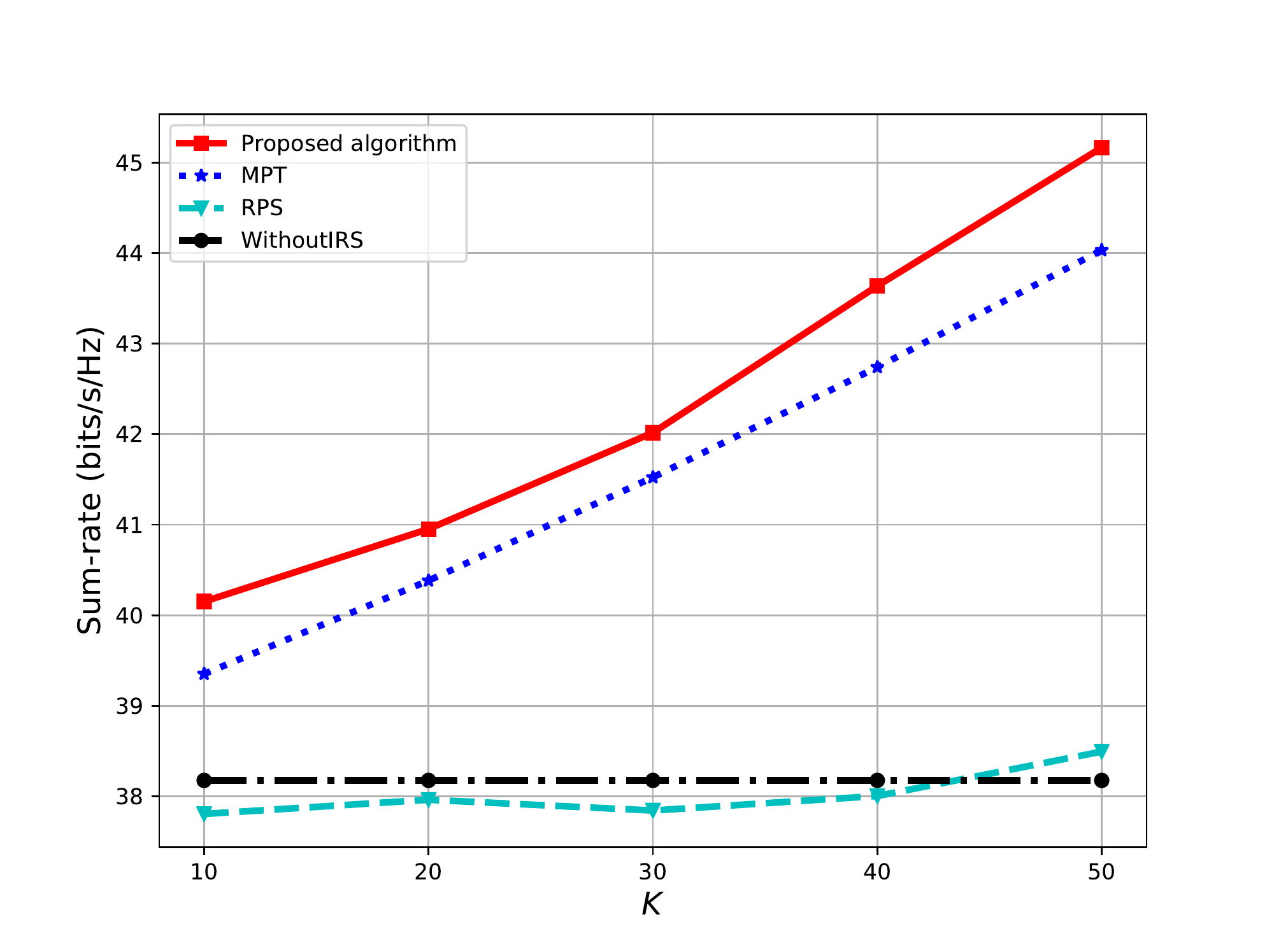}}
	\caption{The network sum-rate versus the number of IRS elements, $K$.}
	\label{fig:NoIRS}
\end{figure}

Firstly, we compare the achievable network sum-rate provided by our proposed algorithm with that of other schemes. Fig. \ref{fig:NoIRS} plots the sum-rate versus different numbers of the IRS elements, $K$, where the number of D2D pairs is set to $N=5$. As can be observed from this figure, the PPO algorithm-based technique outperforms other schemes and is followed by the MPT technique. The RP and WithoutIRS schemes show poorer performance in terms of the network sum-rate. The achievable network sum-rate using our proposed algorithm and MPT improves with increasing the number of IRS elements. The results show that with the monotonic increase in the value of $K$, the communication quality between the D2D-Tx and associated D2D-Rx is enhanced, while the interference from other D2D-Txs is suppressed.

\begin{figure}[h!]
	\centering
	\subfigure{\includegraphics[width=0.45\textwidth]{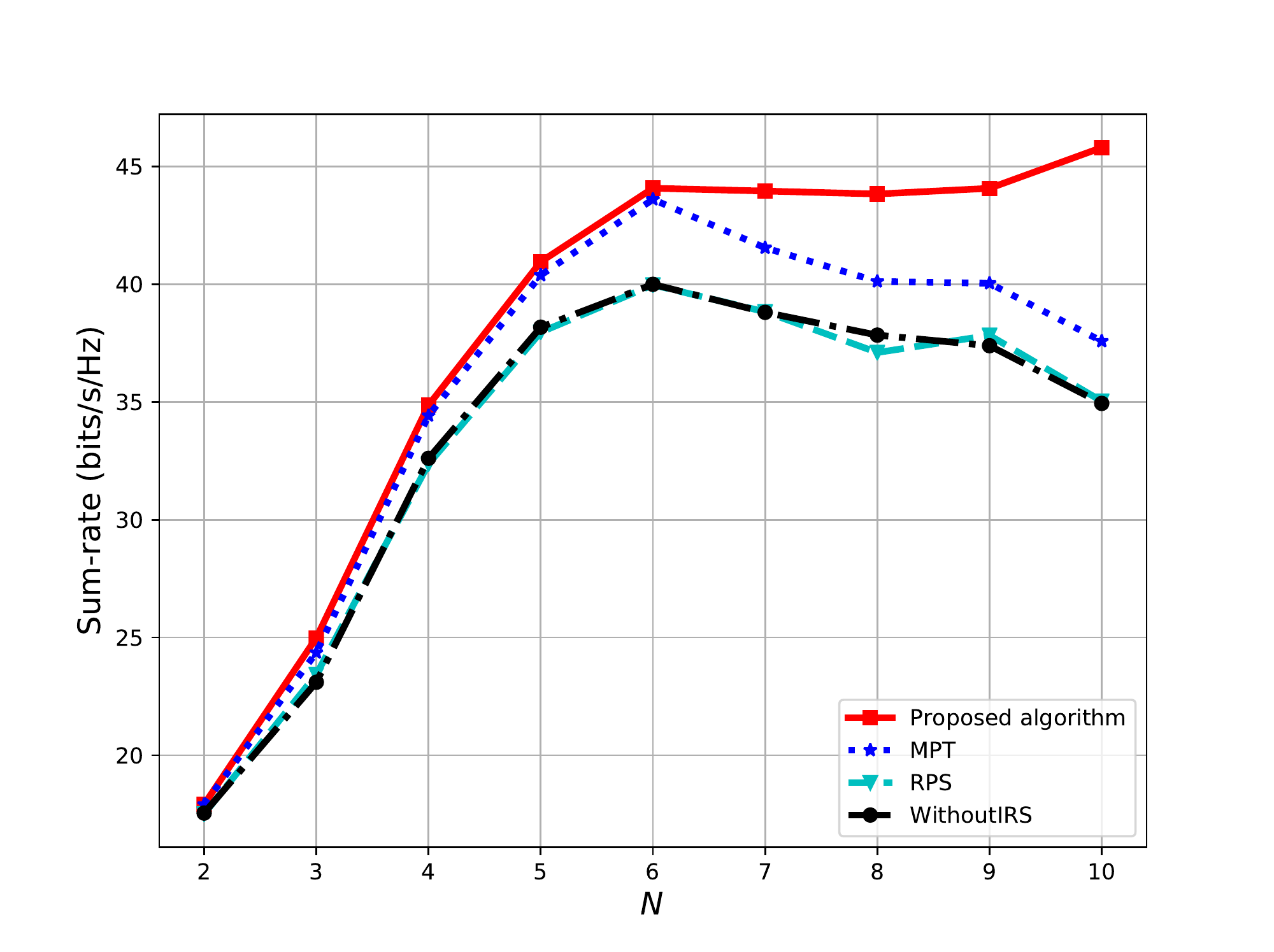}}
	\caption{The network sum-rate versus the number of D2D pairs, $N$.}
	\label{fig:NoD2D}
\end{figure}

Next, the performance of the previously mentioned four schemes is compared while varying the number of D2D pairs, $N$, in Fig. \ref{fig:NoD2D}. We set the number of IRS element to $K=20$ and take the average over $500$ episodes to obtain the results. Our proposed algorithm shows better performance, followed by MPT. With higher number of D2D users, $N \ge 6$, the performance attained by the proposed algorithm still increases while it decreases for the other schemes. The RPS and WithoutIRS models show the worse performance.

\begin{figure}[h!]
	\centering
	\subfigure{\includegraphics[width=0.45\textwidth]{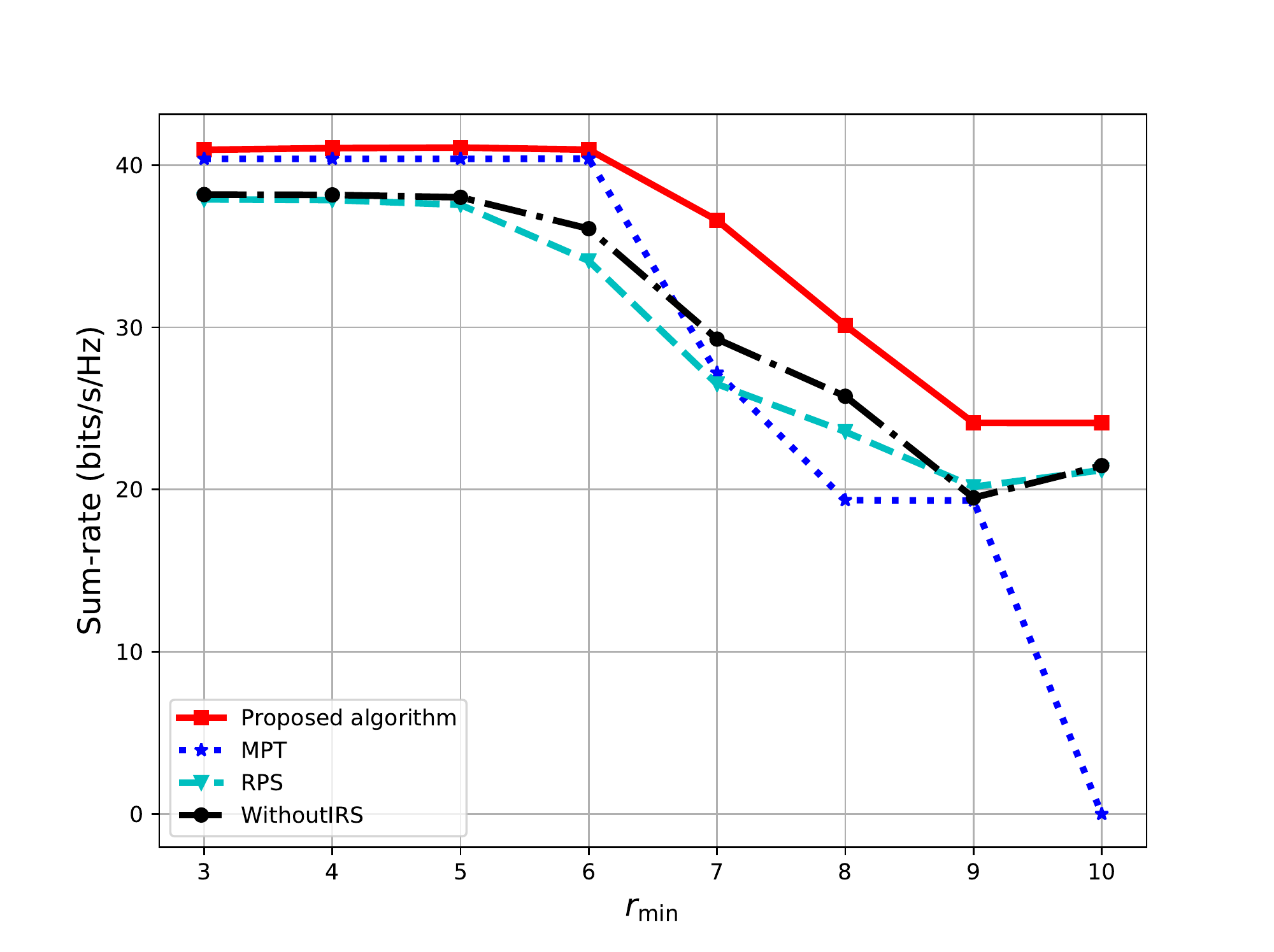}}
	\caption{The network sum-rate versus the QoS threshold, $r_{\min}$.}
	\label{fig:QoS}
\end{figure}

Further, we set $N=5$, $K=20$ and compare the performance results of the four schemes while changing the value of the threshold, $r_{\min}$, in Fig. \ref{fig:QoS}. When the value of $r_{\min}$ increases towards infinity, the number of D2D pairs that satisfies the QoS constraints decreases and the sum-rate of all schemes tends to $0$. The proposed algorithm outperforms the other schemes for all values of $r_{\min}$. The gap between our algorithm and others increases following the increase in $r_{\min}$ when $r_{\min} \ge 5$. The MPT algorithm exhibits the worst performance when $r_{\min} \ge 7$. This suggests that the optimisation of power allocation is important for efficient D2D communications.

\begin{figure}[h!]
	\centering
	\subfigure{\includegraphics[width=0.455\textwidth]{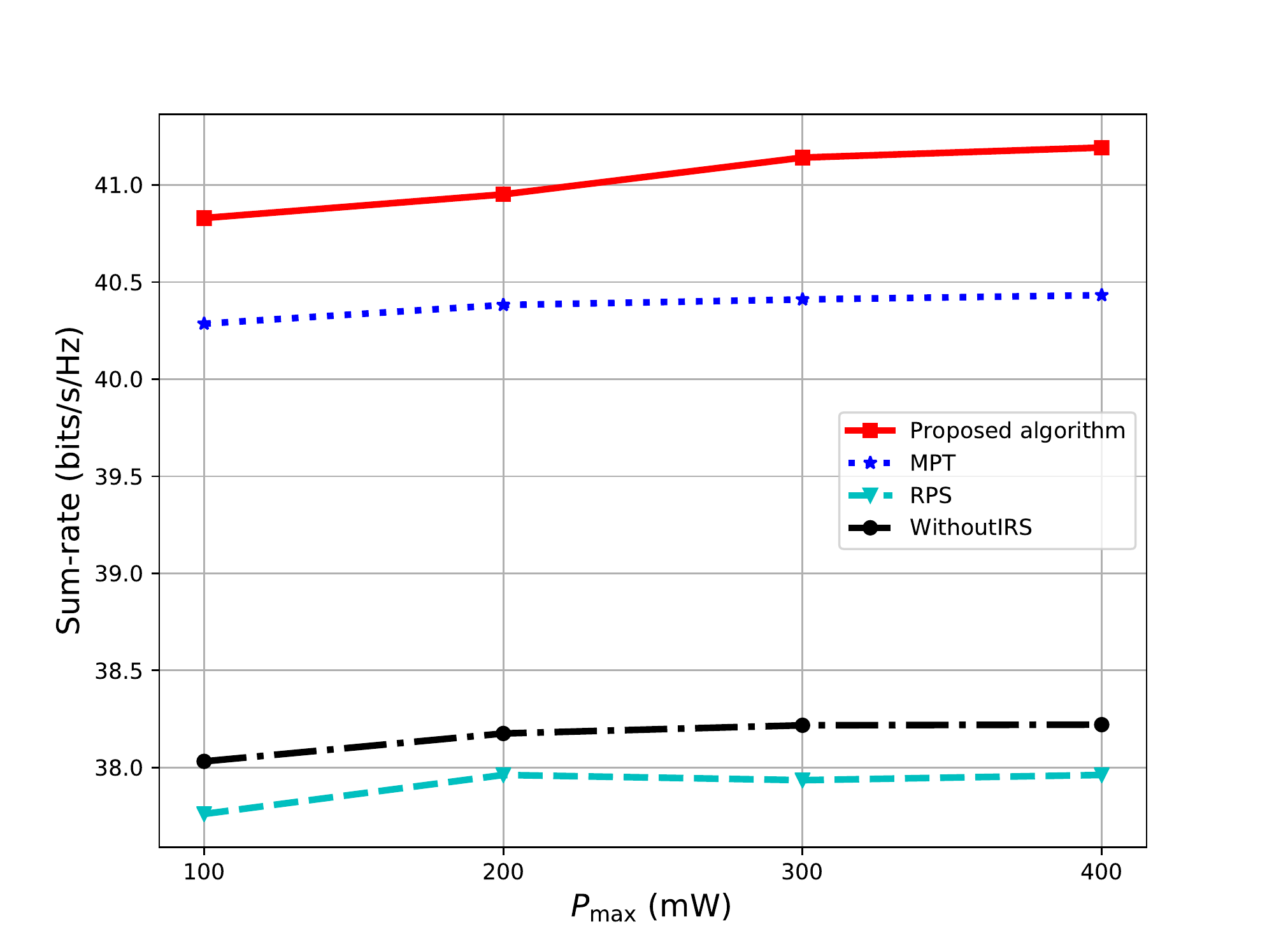}}
	\caption{The network sum-rate versus the maximum transmit power, $P_{\max}$.}
	\label{fig:Pmax}
\end{figure}

Next, we compare the total sum-rate of the four schemes by setting different maximum transmission powers at the D2D-Tx, $P_{\max}$, in Fig. \ref{fig:Pmax}, with $N=5$, $K=20$. As $P_{\max}$ varies from $100$ mW to $400$ mW, the performance of the four schemes increases in the same upward trend. The gap between our proposed algorithm and the other schemes increases with the increase value of $P_{\max}$ as we jointly optimise both power allocation at the D2D-Tx and the IRS's phase shift matrix. It is clear that the proposed algorithm is more effective for mitigating interference and providing a better communication quality.

Furthermore, we use neural networks for establishing the DRL algorithm. Thus, after iterative interactions with the environment, the neural networks are trained for achieving an optimal solution. After training offline, the neural network can be deployed to the system for online execution. The online neural networks can determine the proper action for the IRS phase shift value and the D2D-Tx power allocation for maximising the network sum-rate in real-time.

\section{Conclusion}\label{Sec:Con}
In this paper, we have presented a DRL-based optimal resource allocation scheme for IRS-assisted D2D communications. The PPO algorithm with the clipping surrogate technique has been proposed for joint optimisation of the D2D-Tx power and the IRS's phase shift matrix. Numerical results have showed a significant improvement in the achievable network sum-rate performance compared with the  benchmark schemes. Our proposed scheme demonstrates the superiority of using IRS in mitigating the interference in the D2D communications when compared with other existing schemes.
%

\bibliographystyle{IEEEtran}
\bibliography{IEEEabrv,reference}

\begin{thebibliography}{10}
\providecommand{\url}[1]{#1}
\csname url@samestyle\endcsname
\providecommand{\newblock}{\relax}
\providecommand{\bibinfo}[2]{#2}
\providecommand{\BIBentrySTDinterwordspacing}{\spaceskip=0pt\relax}
\providecommand{\BIBentryALTinterwordstretchfactor}{4}
\providecommand{\BIBentryALTinterwordspacing}{\spaceskip=\fontdimen2\font plus
\BIBentryALTinterwordstretchfactor\fontdimen3\font minus
  \fontdimen4\font\relax}
\providecommand{\BIBforeignlanguage}[2]{{%
\expandafter\ifx\csname l@#1\endcsname\relax
\typeout{** WARNING: IEEEtran.bst: No hyphenation pattern has been}%
\typeout{** loaded for the language `#1'. Using the pattern for}%
\typeout{** the default language instead.}%
\else
\language=\csname l@#1\endcsname
\fi
#2}}
\providecommand{\BIBdecl}{\relax}
\BIBdecl

\bibitem{Khoi:19:Access}
K.~K. Nguyen, T.~Q. Duong, N.~A. Vien, N.-A. Le-Khac, and N.~M. Nguyen,
  ``Non-cooperative energy efficient power allocation game in {D2D}
  communication: A multi-agent deep reinforcement learning approach,''
  \emph{IEEE Access}, vol.~7, pp. 100\,480--100\,490, Jul. 2019.

\bibitem{JH:20:WC}
J.~Huang, C.-C. Xing, and M.~Guizani, ``Power allocation for {D2D}
  communications with {SWIPT},'' \emph{IEEE Trans. Wireless Commun.}, vol.~19,
  no.~4, pp. 2308--2320, Apr. 2020.

\bibitem{HY:20:JSAC}
H.~Yu, H.~D. Tuan, A.~A. Nasir, T.~Q. Duong, and H.~V. Poor, ``Joint design of
  reconfigurable intelligent surfaces and transmit beamforming under proper and
  improper {Gaussian} signaling,'' \emph{IEEE J. Select. Areas Commun.},
  vol.~38, no.~11, pp. 2589--2603, Nov. 2020.

\bibitem{YZ:20:VT}
Y.~Zou, S.~Gong, J.~Xu, W.~Cheng, D.~T. Hoang, and D.~Niyato, ``Wireless
  powered intelligent reflecting surfaces for enhancing wireless
  communications,'' \emph{{IEEE} Trans. Veh. Technol.}, vol.~69, no.~10, pp.
  12\,369--12\,373, Oct. 2020.

\bibitem{BZ:21:TCOM}
B.~Zheng, C.~You, and R.~Zhang, ``Efficient channel estimation for double-{IRS}
  aided multi-user {MIMO} system,'' \emph{IEEE Trans. Commun.}, vol.~69, no.~6,
  pp. 3818--3832, Jun. 2021.

\bibitem{KK:21:NCE}
\BIBentryALTinterwordspacing
K.~K. Nguyen, S.~Khosravirad, L.~D. Nguyen, T.~T. Nguyen, and T.~Q. Duong,
  ``Intelligent reconfigurable surface-assisted multi-{UAV} networks: Efficient
  resource allocation with deep reinforcement learning,'' 2021. [Online].
  Available: \url{https://arxiv.org/abs/2105.14142}
\BIBentrySTDinterwordspacing

\bibitem{YC:21:WC}
Y.~Chen, B.~Ai, H.~Zhang, Y.~Niu, L.~Song, Z.~Han, and H.~V. Poor,
  ``Reconfigurable intelligent surface assisted device-to-device
  communications,'' \emph{IEEE Trans. Wireless Commun.}, vol.~20, no.~5, pp.
  2792--2804, May 2021.

\bibitem{SJ:21:WCL}
S.~Jia, X.~Yuan, and Y.-C. Liang, ``Reconfigurable intelligent surfaces for
  energy efficiency in {D2D} communication network,'' \emph{IEEE Wireless
  Commun. Lett.}, vol.~10, no.~3, pp. 683--687, Mar. 2021.

\bibitem{KK:19:Access}
K.~K. Nguyen, T.~Q. Duong, N.~A. Vien, N.-A. Le-Khac, and L.~D. Nguyen,
  ``Distributed deep deterministic policy gradient for power allocation control
  in {D2D}-based {V2V} communications,'' \emph{IEEE Access}, vol.~7, pp.
  164\,533--164\,543, Nov. 2019.

\bibitem{Khoi:20:Access}
K.~K. Nguyen, N.~A. Vien, L.~D. Nguyen, M.-T. Le, L.~Hanzo, and T.~Q. Duong,
  ``Real-time energy harvesting aided scheduling in {UAV}-assisted {D2D}
  networks relying on deep reinforcement learning,'' \emph{IEEE Access},
  vol.~9, pp. 3638--3648, Dec. 2021.

\bibitem{CH:20:JSAC}
C.~Huang, R.~Mo, and C.~Yuen, ``Reconfigurable intelligent surface assisted
  multiuser {MISO} systems exploiting deep reinforcement learning,'' \emph{IEEE
  J. Select. Areas Commun.}, vol.~38, no.~8, pp. 1839--1850, Aug. 2020.

\bibitem{MS:21:VT}
M.~Shokry, M.~Elhattab, C.~Assi, S.~Sharafeddine, and A.~Ghrayeb, ``Optimizing
  age of information through aerial reconfigurable intelligent surfaces: A deep
  reinforcement learning approach,'' \emph{{IEEE} Trans. Veh. Technol.},
  vol.~70, no.~4, pp. 3978--3983, Apr. 2021.

\bibitem{KF:20:WCL}
K.~Feng, Q.~Wang, X.~Li, and C.-K. Wen, ``Deep reinforcement learning based
  intelligent reflecting surface optimization for {MISO} communication
  systems,'' \emph{IEEE Wireless Commun. Lett.}, vol.~9, no.~5, pp. 745--749,
  May 2020.

\bibitem{KK:21:TCOM}
\BIBentryALTinterwordspacing
K.~K. Nguyen, T.~Q. Duong, T.~Do-Duy, H.~Claussen, and L.~Hanzo, ``{3D UAV}
  trajectory and data collection optimisation via deep reinforcement
  learning,'' 2021. [Online]. Available: \url{https://arxiv.org/abs/2106.03129}
\BIBentrySTDinterwordspacing

\bibitem{BD:95:Book:v1}
D.~P. Bertsekas, \emph{{Dynamic Programming and Optimal Control}}.\hskip 1em
  plus 0.5em minus 0.4em\relax Athena Scientific Belmont, MA, 1995, vol.~1,
  no.~2.

\bibitem{JS:17:PPO}
\BIBentryALTinterwordspacing
J.~Schulman, F.~Wolski, P.~Dhariwal, A.~Radford, and O.~Klimov, ``Proximal
  policy optimization algorithms,'' 2017. [Online]. Available:
  \url{https://arxiv.org/abs/1707.06347}
\BIBentrySTDinterwordspacing

\bibitem{JS:16:ICLR}
J.~Schulman, P.~Moritz, S.~Levine, M.~I. Jordan, and P.~Abbeel,
  ``High-dimensional continuous control using generalized advantage
  estimation,'' in \emph{Proc. 4th International Conf. Learning Representations
  (ICLR)}, 2016.

\bibitem{Mnih:16}
V.~Mnih \emph{et~al.}, ``Asynchronous methods for deep reinforcement
  learning,'' in \emph{Proc. Int. Conf. Mach. Learn.}\hskip 1em plus 0.5em
  minus 0.4em\relax PMLR, 2016, pp. 1928--1937.

\bibitem{Abadi:16}
M.~Abadi \emph{et~al.}, ``{Tensorflow}: A system for large-scale machine
  learning,'' in \emph{Proc. 12th USENIX Sym. Opr. Syst. Design and Imp. (OSDI
  16)}, Nov. 2016, pp. 265--283.

\end{thebibliography}

\end{document}